# Prospects towards Paired Electrolysis at Industrial Currents

Lu Xia[1,#], Kaiqi Zhao[1,#], Sunil Kadam[1], M. Dolores Blanco-González[2], María D. Hernández Alonso[2], F. Pelayo García de Arquer[1,*]

[1]ICFO - Institute de Ciències Fotòniques, The Barcelona Institute of Science and Technology, Castelldefels (Barcelona), 08860, Spain

[2]REPSOL S.A. - Repsol Technology Lab, C/ Agustín de Betancourt s/n, Móstoles (Madrid) 28935, Spain

[*]corresponding author: pelayo.garciadearquer@icfo.eu
[#]These authors contribute equally.

## Abstract

**Paired electrolysis at industrial current densities offers an energy-efficient and sustainable alternative to thermocatalytic chemical synthesis by leveraging anodic and cathodic valorization. However, its industrial feasibility remains constrained by system integration, including reactor assembly, asymmetric electron transfer kinetics, membrane selection, mass transport limitations, and techno-economic bottlenecks. Addressing these challenges requires an engineering-driven approach that integrates reactor architecture, electrode-electrolyte interactions, reaction pairing, and process optimization. Here, we discuss scale-specific electrochemical reactor assembly strategies, transitioning from half-cell research to full-scale stack validation. We develop reaction pairing frameworks that align electrocatalyst design with electrochemical kinetics, enhancing efficiency and selectivity under industrial operating conditions. We also establish application-dependent key performance indicators (KPIs) and benchmark propylene oxidation coupled with hydrogen evolution reaction (HER) or oxygen reduction reaction (ORR) against existing industrial routes to evaluate process viability. Finally, we propose hybrid integration models that embed paired electrolysis into existing industrial workflows, overcoming adoption barriers.**

# Introduction

The continued reliance on fossil fuels for chemicals and fuels raises environmental and economic concerns, highlighting the need for renewable energy-powered sustainable processes.[1–3] The rapid deployment of renewable energies led to a global 3.4 TW installed capacity by 2022.[4] Storing renewable energy in the form of chemical bonds would offer a path to synthesize such chemicals and buffer the intermittent character of renewables.[5]

Water electrolysis is a key strategy for generating clean hydrogen,[6] relying on coupled redox reactions where water ($H_2O$) or protons ($H^+$) are reduced to $H_2$ at the cathode, while water or hydroxyl groups ($OH^-$) are oxidized to $O_2$ via the oxygen evolution reaction (OER) at the anode.[7] Beyond hydrogen production, similar electrolysis processes can be applied to carbon- and nitrogen-based waste molecules, enabling sustainable chemical synthesis through tailored redox reactions. The efficiency of electrolysis hinges on achieving high activity at low overpotentials, with the OER presenting the greatest thermodynamic and kinetic challenges.[8,9]

The limited economic value of $O_2$ (\$24–40/ton $O_2$),[10] combined with the high overpotentials required for the OER, has motivated the search for alternative anodic reactions. These reactions can substitute the OER by producing higher-value products, while addressing its kinetic and thermodynamic limitations.

Anodic half reactions provide a versatile platform for converting waste or low-value reactants into higher-value products through organic electrosynthesis,[11] plastic upcycling, biomass valorization, and commodity chemical oxidation.[12] For example it can upcycle plastics,[13] by breaking down polymers like polyethylene terephthalate (PET), polybutylene terephthalate (PBT), polyethylene (PE), polyvinyl chloride (PVC), polystyrene (PS) into reusable feedstocks. Additionally, anodic reactions can valorize complex biomass materials,[14] such as converting hydroxymethyl-furfural (HMF) into 2,5-Furan-dicarboxylic acid (FDCA).[15]

Compared to conventional thermocatalytic processes, electrochemical reactions operate under milder conditions, reducing energy consumption and safety risks.[16] For example, the synthesis of ethylene and propylene oxides traditionally requires 200–300°C and 1–3 MPa, while electrochemical reactions can achieve similar outcomes under atmospheric conditions. Furthermore, electrochemical reactions facilitate coupling reactions (*e.g.*, C–N,[17] N–H,[18] N–N,[19] bonds), broadening its industrial relevance.

Paired electrolysis offers a sustainable route towards complete electron economy for simultaneous anodic and cathodic valorization. However, its industrial implementation faces critical challenges: *i)* At scale, increasing interface complexity disrupts mass transport in reactors, while membrane stability and ion crossover affect efficiency and product purity. *ii)* The inherent asymmetry in electron transfer rates and overpotentials between anodic and cathodic reactions leads to current imbalances and energy losses. *iii)* Reaction-specific KPIs and downstream separation complicate performance evaluation, requiring careful management of trade-offs in selectivity, energy efficiency, and operational stability. For successful scaling, paired electrolysis must be both *iv)* economically competitive and *v)* easily integrated into existing industrial infrastructure (**Fig. 1a**).

Here, we present an engineering-driven framework for advancing paired electrolysis from laboratory feasibility to industrial deployment (**Fig. 1b**). We introduce a scale-specific reactor integration strategy, transitioning from half-cell studies to full-cell and industrial-scale implementations to address mass transport, interfacial challenges, and integration bottlenecks. We develop reaction pairing strategies that optimize electron transfer kinetics. To establish industrial viability, we define application-dependent KPIs and benchmark propylene oxidation paired with HER or ORR as a case study against high-TRL thermocatalytic routes. Finally, we propose hybrid integration strategies to embed paired electrolysis into existing industrial workflows, facilitating scalability and techno-economic feasibility.

**Reactor design and implementations**

Reactor design is a cornerstone of advancing alternative paired electrolysis, bridging the gap between fundamental research and industrial application. Thoughtfully engineered reactors not only enhance KPIs, such as current density, selectivity, and energy efficiency but also address critical scalability challenges. Below, we outline a TRL-specific reactor design strategy, detailing the transition from half-cell studies to full-scale electrolyzers, ensuring both scientific insight and practical feasibility.

**Half or flow cells for early research stages (low TRLs)**

Initial cathodic or anodic research often uses half-cell setups like split H-cells, unsplit cells, and half-flow cells (**Fig. 2a**). Split H-cells suit gas-producing reactions like chlorine evolution, while unsplit cells, with reference electrodes, allow detailed mechanistic studies.

The reactor design plays a crucial role in directing selectivity. For example, in N–heteroarenes C–H carboxylation: *i)* Split cell favors $C_5$–carboxylation due to a highly reducing cathodic environment, where sequential electron transfer facilitates carboxylation, followed by oxidation by trace oxygen in the system. *ii)* Unsplit cell promotes $C_4$–carboxylation as anodic oxidation generates hydrogen acceptors, facilitating hydrogen atom transfer (HAT) or proton-coupled electron transfer (PCET) processes that lower the energy barrier for $C_4$ functionalization.[20] This example illustrates how reactor design governs selectivity by controlling redox environment, mass transport, and interfacial effects, optimizing electrochemical valorization.

Three-electrode half-flow cells offer enhanced mass transport, separation of compartments, and precise control over electrode potential, making them versatile for various electrochemical reactions.[21] The use of flow reactors has shown distinct behavior compared to static reactors, revealing additional radical adducts beyond the typical •OH species. This further highlights the impact of reactor and mass-transport in promoting different reaction intermediates, pathways, and mechanisms.[22]

Such non-zero-gap cells, while limited by higher internal resistance and large polar distances, offer better separation and flexibility, making them suitable for early-stage paired electrolysis studies, particularly for alcohol oxidation paired with HER,[23] a common benchmark reaction due to its well-understood chemistry and lower activation energy requirements.[24] Although these cells are commonly applied to liquid reactants like alcohols, their flexibility also provides potential for other oxidation reactions involving gas molecules, including methane,[25] ethylene, propylene,[16] and aldehyde.[26]

**Full electrolyzers for intermediate research stages (Medium TRLs)**

As TRL increases (TRLs 4–6), reactor designs prime scalability, voltage efficiency, and reliability. Zero-gap reactors optimize voltage losses and mass transport (**Fig. 2b**) by placing anode and cathode electrodes together with a membrane, in the absence of dedicated catholyte and anolyte chambers.

In these systems, beyond separation and ion transport, membranes critically influence water availability and the electrochemical reaction environment. Common membranes (**Fig. 2d**) for water electrolysis include diaphragm separators,[27] proton exchange membranes (PEMs),[28] anion exchange membranes (AEMs),[29] bipolar membranes (BPMs),[30] and ion-solvating membranes (ISMs).[31] However, many electrochemical reactions have been designed to operate in neutral environments,[32] which stand in contrast to the well-established acidic or alkaline conditions used in water electrolysis. Membranes tailored for neutral pH operation face distinct challenges compared to acidic or basic environments, including reduced ionic conductivity and heightened contamination risks. Unlike in strong acid or base systems, ion transport is less efficient in neutral electrolytes, and side reactions can further compromise performance. Therefore, these membranes must achieve a careful balance between efficient ion transport, contaminant rejection, and stability to ensure selective operation under these unique conditions.[33]

**Stack for industrial application stages (High TRLs)**

At higher TRLs (TRL 7–9), the emphasis is on commercialization, which involves further scaling up and integrating the process into industrial settings (**Fig. 2c**), like high current density (>1 A/cm$^2$). (Ref.[34]) At this stage, reactor designs prioritize economic viability and process reliability, leveraging large flow systems and high-surface-area electrodes to achieve industrial-scale operation. Full process integration, and eventually membrane-free operation, may offer a path to reduced energy intensity and costs. For example, selective oxidation of glycerol and ethylene glycol was achieved in a 150 cm$^2$ membrane-free electrolyzer, with high-rate production of lactic acid and glycolic acid at 56.9 and 36.8 mmol/h, respectively.[35] At high TRLs, the energy intensity of the full process—including product management, operational stability, and safety—becomes a dominant factor, surpassing the importance of individual reaction KPIs.

**TRL-driven tasks & milestones: from concept to commercialization**

Paired electrolysis follows a TRL-driven roadmap (**Fig. 2e**) from lab to industry. Early-stage research (TRL 1–3) focuses on active sites and reaction pathways (*e.g.*, half-cell chiral synthesis, plastic upcycling etc.). Prototype demonstration (TRL 4–6) advances electrode/membrane engineering and small-stack testing (*e.g.*, HMF to FDCA paired with HER ORR *etc.*). Industrial deployment (TRL 7–9) involves large-scale validation, separation, and techno-economic analysis (*e.g.*, chlor-alkali electrolysis). This structured approach ensures scalability and process integration.

**Contribution of reactor design to effective paired electrolysis systems**

Reactor design strengthens electrolysis efficiency by reducing side reactions, minimizing energy losses, and enabling scalable operation from lab to industry. Paired electrolysis systems operate in chemically diverse environments, often leading to membrane fouling, side reactions, and degradation that reduce stability and selectivity. For instance, organic electrolytes or redox-active species can interact with the membrane, leading to fouling or degradation that undermines stability and selectivity. Surface-modified membranes (*e.g.*, Perfluorosulfonic acid with graphene oxide) enhance ion selectivity, resist fouling, and prevent degradation, making them well-suited for paired electrolysis.[36,37]

Moreover, functionalized nanomaterials improve membrane resistance to organic electrolytes, preventing swelling or degradation in the presence of aggressive solvents. Fouling is also mitigated by incorporating antifouling agents or catalytic nanoparticles that resist the buildup of organic compounds on the membrane surface.

In gaseous reaction, managing gas-liquid interfaces is critical to prevent gas crossover and maintain selectivity, often addressed using hydrophobic channels or gas-diffusion layers.

## Challenges & opportunities for industrially viable paired electrolysis

Industrially viable paired electrolysis requires careful pairing with cathodic reactions to ensure charge balance and optimize reaction efficiency. The selection of reaction pairs is critical, as it directly influences system scalability, economic viability, and environmental impact.

**Availability of reactants**

Anodic reaction can be effectively paired with cathodic hydrogen generation or $CO_2$ reduction, creating valuable opportunities for co-valorization across various sectors. On the cathodic side, common reactants like $H_2O$, $CO_2$, and $N_2$ are appealing due to their availability in waste streams and atmospheric abundance, with processes like hydrogen evolution reaction (HER),[38–40] oxygen reduction reaction (ORR),[41,42] $CO_2$ reduction ($CO_2$R),[43,44] and nitrogen/nitrate reduction ($N_2$R, $NO_x$R) being widely studied (**Fig. 3a**). On the anodic side, potential reactants include readily available low-cost feedstocks such as chloride ions from seawater,[45,46] and biomass derivatives (*e.g.*, HMF, lignocellulosic materials).[47] High-value reactions like HMF to FDCA demonstrate significant economic potential, while moderate value conversions, such as glycerol to acrylic acid, offer a balance between scalability and profitability (**Fig. 3b**).[48]

Additionally, moderate-volume reactions like ethylene glycol to glycolic acid and 1,2-propanediol to lactic acid offer a balanced opportunity, with both market demand and notable price increases. These could appeal to specialty chemical markets where the benefits, like enhanced selectivity and sustainability, add further value.

For high-volume, lower-margin reactions, such as methanol to formaldehyde and ethanol to acetic acid, anodic reactions would need to be cost-competitive with existing production methods. However, the potential environmental benefits of electrochemical reactions (combining waste valorization and the use of low carbon footprint energy) could drive its adoption, especially for reactions aligning with sustainability goals. For example, producing 1 kg of FDCA emits 46.58 kg of $CO_2$ equivalent via thermochemical methods, but only 28.01 kg via the electrochemical route, achieving a 40% reduction.[49]

Paired electrolysis and anodic valorization share the use of anodic and cathodic reactions but differ in objectives. While paired electrolysis maximizes system efficiency by co-valorizing both reactions, anodic valorization focuses on replacing OER with value-added anodic reactions, often leaving the cathodic reaction secondary (*e.g.*, ORR).

Anodic valorization can be considered a specialized form of paired electrolysis when coupled with HER, where the anodic reaction is designed for high-value product generation while sustaining efficient hydrogen production at the cathode.

In practical systems, optimizing paired electrolysis is crucial to balance anodic and cathodic reactions, ensuring both contribute to overall system efficiency and value. This dual-focus strategy maximizes electron economy and aligns with industrial and sustainability objectives.[50]

**Thermodynamics: standard and operational potential**

Anodic reactions involve both direct and indirect oxidation pathways, depending on the nature of the reactants and intermediates. In indirect oxidation, inorganic intermediates, such as hydroxyl radicals, are preferentially oxidized and subsequently facilitate the oxidation of organic compounds. This mechanism is particularly prevalent in organic reactions, enabling the conversion of complex molecules through highly reactive intermediates (**Fig. 3c**).[51]

The standard oxidation potential of most small molecules, such as alcohols and amines, is lower than that of the OER.[52] However, slow reaction kinetics and suboptimal catalyst surface properties often necessitate higher anodic potentials (**Fig. 3d**) to achieve practical reaction rates.

For higher-potential reactions, including $Cl^-$ to $Cl_2$ and $SO_3^{2-}$ to $SO_4^{2-}$, achieving voltage savings remains a challenge due to their substantial thermodynamic energy demands. Even with advanced catalysts, only modest reductions in operating voltage may be attainable, though these can still contribute to making the process more feasible and economically viable. This challenge also applies to multi-electron reactions, such as furfural to FDCA and glycerol oxidation.

When the current density increases to industrial level benchmarks for HER (*e.g.*, 1 A/cm$^2$), the anodic electrode potential can escalate up to as high as 3–5 V (vs. RHE). In such cases, the voltage of the entire electrolyzer can reach up to 6–8 V.[53]

The high voltages observed in electrolysis, compared to theoretical potential differences, are driven by both thermodynamic and kinetic factors. Thermodynamically, the large number of electron transfers required in complex organic transformations increases energy demands, while kinetically, barriers to bond breaking and formation necessitate higher potentials to achieve practical rates.

Catalyst optimization, including tailored surface area, morphology, and active site distribution, is crucial for overcoming kinetic barriers.[54] Functional group modifications, such as hydroxyl (–OH), carboxyl (–COOH), and sulfonic acid (–SO$_3$H),[55,56] further enhance reactivity by enabling selective interactions (*e.g.*, hydrogen bonding) with organic reactants, reducing overpotentials and improving selectivity.

From a system perspective, ohmic losses, caused by resistance in the electrolyte, electrodes, and interconnects, increase voltage requirements, compounding thermodynamic and kinetic inefficiencies. Optimizing electrolyte composition and membrane conductivity is critical to minimizing these losses, particularly in paired electrolysis systems where organic reactants dominate ion transport processes.

In water splitting, protons (H$^+$) or hydroxide ions (OH$^-$) are both the charge carriers and the main reactants for the HER and OER. In half electrolysis reaction, while H$^+$ or OH$^-$ may still act as charge carriers to facilitate ion transport, they are not the main reactants. Instead, the key reactants are typically organic molecules or other substrates. This means the focus shifts to optimizing the electrolyte and membrane to reduce ohmic losses, as the system relies more on efficient ion transport rather than the direct involvement of H$^+$ or OH$^-$ in the reactions.

Mass transport limitations hinder energy efficiency due to the slow diffusion of larger organic molecules, causing concentration polarization and higher voltage needs. Hierarchical porous electrode designs can address this by improving reactant diffusion and product removal.

**Market prices and volume of products**

The economic viability hinges on market dynamics, particularly the prices and demand of target products. A clear understanding of these factors is essential to assess its competitiveness and guide system optimization.

Cathodic products (**Fig. 3a**), like green hydrogen, range in price from $1.39 to $15/kg,[52,57] depending on production methods, while ethylene and propylene, priced at $1.0–1.1/kg,[58] have high global demands of 162 Mt/year and 11.9 Mt/year in 2022,

respectively, with projected growth exceeding 4% annually.[59,60]

On the anodic side, products vary in value and volume. For example, 2,5-furandicarboxylic acid (FDCA), priced at $1.2–4.0/kg, has a modest output of 0.5 Mt/year,[52,61,62] while ethylene and propylene oxides ($1.3–1.7/kg) are crucial for producing polyols used in polyurethane manufacturing.[63,64]

Balancing product price and output is key to optimizing paired electrolysis. High-value, low-volume products like dihydroxyacetone ($2.0/kg, 0.004 Mt/year) suit niche markets, while bulk chemicals like formaldehyde ($0.555/kg, 18 Mt/year) align with large-scale production (**Fig. 3b**). Pairing scalable cathodic reactions with high-margin anodic reactions maximizes economic potential.

**Number of electrons and rate-determining step**

The number of electrons transferred in a reaction influences the complexity of the mechanism but is not the sole determinant of kinetics; the rate-determining step (RDS) ultimately governs the overall reaction rate. For example, in the six-electron oxidation of HMF to FDCA, multiple intermediates and reaction steps complicate the identification and optimization of the RDS. However, electron economy can still be achieved with high Faradaic efficiency if catalysts and reaction conditions are properly optimized. Reducing the number of electron transfers may lower overpotentials in some anodic reactions, but multi-electron processes, such as HMF oxidation, often remain necessary to achieve desired products, underscoring the need for tailored catalyst design and mechanistic understanding.

**Side effects of paired electrolysis**

Several challenges must be addressed to ensure the success of paired electrolysis in industrial applications (**Table 1**):

*i)* **Crossover and contamination:** Product and anion crossover reduce purity and yield. Optimizing selective catalysts, membranes, and ion-selective electrodes can mitigate these effects and enhance product quality.

*ii)* **Surface site competition:** Competing species at electrode surfaces reduce efficiency. Designing selective catalysts and tailored electrolytes can optimize surface utilization, improving reaction selectivity and energy efficiency.

*iii)* **Electrolyte and pH mismatch:** Different anodic and cathodic reactions often require incompatible electrolytes or pH conditions, leading to fouling or inefficiencies. Developing dual-compatible electrolyte systems or pH buffers is essential for integrated paired electrolysis systems.

*iv)* **Market demand mismatch:** Discrepancies in market demand for anodic and cathodic products can hinder large-scale adoption. Pairing anodic reactions with scalable cathodic reactions, such as hydrogen evolution, can better align production volumes and market needs.

*v)* **Diverse application requirements:** Paired electrolysis spans varied applications, from biomass valorization to fine chemical production, each with distinct energy and selectivity demands. Customized solutions, including advanced catalysts and reactor designs, are critical to addressing these challenges.

**Table 1** | Comparison of different anodic pairing processes: challenges, key requirements and applications.

| Processes (scale) | Challenges/requirements | Key performance indicators (KPIs) |
|---|---|---|
| Biomass valorization (Large-scale, industrial)[35,47,65] | <ul><li>Scalability</li><li>Feedstock variability</li><li>High energy demands</li><li>Requires high throughput</li><li>Low cost</li><li>Sustainable feedstock use</li></ul> | <ul><li>Current density</li><li>Faradaic & energy efficiency</li><li>Stability (long-term operation for industrial scale)</li><li>Product yield (ensuring economic viability)</li></ul> |
| Ethylene/propylene oxidation (Commodity-level, industrial)[66–68] | <ul><li>High energy demands</li><li>Selectivity</li><li>Environmental impact</li><li>Requires efficient</li><li>Reduced costs</li></ul> | <ul><li>Selectivity (minimize side products)</li><li>Energy consumption (key to reducing operational costs)</li><li>Faradaic efficiency (maximize output with input electricity)</li><li>Current density (important for large-scale operation)</li></ul> |
| Organic electrosynthesis (Small-scale, high-value)[69–71] | <ul><li>High selectivity</li><li>Product purity</li><li>Regulatory compliance</li><li>Requires precise</li></ul> | <ul><li>Selectivity (critical for precision in high-value products)</li><li>Product purity (essential for pharmaceuticals and fine chemicals)</li></ul> |

| | reaction control, selective catalysts, and high purity | • Stability (ensuring consistent production over time)<br>• Faradaic/energy efficiency (less critical than purity/selectivity) |
|---|---|---|
| Plastic upcycling (Large-scale or batch)[72–74] | • Degradation control<br>• Polymer variability<br>• Process efficiency<br>• Requires effective depolymerization, low energy input, and minimal contamination | • Depolymerization yield (key to breaking down polymers effectively)<br>• Energy consumption (critical for keeping the process economical)<br>• Product yield (important for the economics of upcycling)<br>• Selectivity (minimizing unwanted byproducts) |

**Downstream separation**

A major challenge in paired electrolysis is the separation of supporting electrolytes from target products, which increases operational costs and limits scalability in industrial applications. Recent advances, such as selective membranes and solid electrolytes, may help address this issue by enabling targeted ion or product transport while minimizing contamination.[75,76]

Solid electrolytes enhance downstream separation in paired electrolysis by selectively transporting target ions while blocking contaminants. This reduces product contamination, simplifies separation processes, and improves scalability and cost-efficiency. For example, solid oxide electrolytes have demonstrated high selectivity in the electrooxidation of biomass-derived molecules like glycerol, directly reducing downstream separation costs by preventing byproduct crossover and enabling the efficient recovery of high-purity products. This concept could be extended to other anodic processes, such as propylene oxidation, where solid electrolytes can streamline the separation of oxidation products.[77–80]

Additionally, designing processes that minimize energy use in separation steps, such as liquid-to-liquid upgrading, improves the scalability by simplifying product separation steps, enhancing product purity and reducing energy demands.[33] It simplifies downstream processes with easier product recovery, lower energy use, and better mass transfer, avoiding energy-intensive steps like condensation in gas-to-liquid systems.

**Key performance indicators (KPIs)**

KPIs define the potential of paired electrolysis to achieve sustainable chemical valorization at scale. Small-scale research on half-cells often demonstrates selectivity exceeding 80% at current densities of 10 mA/cm$^2$. However, these results are far from meeting industrial standards, such as PEM and AEM systems operating at current densities ≥1 A/cm$^2$.

Recent advancements in high-current applications **(Fig. 4a)** have achieved Faradaic efficiencies (FE) between 60% and 96% at 0.2–2.0 A/cm$^2$. However, each anodic pairing process has different challenges and requires tailored research to meet specific KPIs (**Table 1**).

In biomass valorization, achieving high current densities is essential for processing large feedstock volumes efficiently. Selective catalysts and scalable reactors must be designed to handle the variability of biomass while maintaining industrial-level current densities. For instance, an Au/Ni(OH)$_2$ catalyst achieved selective electrooxidation of glycerol and ethylene glycol into lactic acid (77%) and glycolic acid (91%) at > 0.3 A/cm$^2$, demonstrating industrial relevance.[35]

For organic electrosynthesis in chemical production, selectivity and product purity are key,[69-71, 81] but long-term stability is equally critical for industrial applications. Research must focus on highly selective catalysts and precise reaction control to ensure consistent, high-quality outputs over extended operational periods. For example, a recent study on cobalt-catalyzed enantioselective C–H activation demonstrated high selectivity in synthesizing chiral compounds, achieving up to 99% enantiomeric purity while coupling with hydrogen evolution.[69]

Commodity-level anodic upgrading focuses on high FE, selectivity for valuable products, industrial-level current densities (>1 A/cm$^2$), and long-term stability. Optimizing catalysts and minimizing energy use are key for scalability and cost-effectiveness. For example, a barium-oxide-loaded catalyst was designed to suppress the deactivation of active species, achieving a FE of 90% for Cl$_2$-mediated ethylene oxidation to ethylene oxide.[82]

Plastic upcycling demands efficient systems with high depolymerization rates. For instance, a Pd–Ni foam catalyst converted PET with 100% efficiency at 400 mA/cm² and 0.7 V vs. RHE, yielding 99% terephthalate and 93% FE for carbonate, while producing hydrogen.[83] Another Pd–Ni(OH)$_2$/Ni foam catalyst transformed PET-derived ethylene glycol into glycolic acid with 85% FE, 90% selectivity, and over 200 hours of stability at 0.6 A/cm² and 1.15 V vs. RHE.[84]

Other anodic reactions prioritize energy efficiency while ensuring high current density stability. For instance, ascorbic acid electrooxidation reached 2 A/cm$^2$ at 1.1 V with nearly 100% efficiency, consuming just 2.63 kWh/Nm$^3$. (ref.[84]) Similarly, HMF oxidation achieved a 200% FE at 100 mA/cm$^2$, using only 0.35 kWh/Nm$^3$—14 times less than conventional water electrolysis.[26]

**Competitive landscape of paired electrolysis**

Beyond meeting KPIs, paired electrolysis must establish a clear competitive advantage over alternative decarbonization methods to gain market acceptance. For example, alternative decarbonization routes for the production of polyols includes *i)* cracker electrification (e-cracker) with renewable electricity, *ii)* bio-polyols from vegetable oils or lignin (bio-based polyols), *iii)* $CO_2$-based polyols, processes in which $CO_2$ is incorporated in the polyol as a monomer (Econic, Covestro, Aramco, Repsol have developments on these $CO_2$-based polyols), *iv)* recycling of polyurethane for polyol production (**Fig. 4b**).

These alternative decarbonization routes vary in feasibility and scalability. For instance, bio-based polyols reduce reliance on petrochemical feedstocks but face challenges in raw material availability. $CO_2$-based polyols directly address carbon emissions by incorporating $CO_2$ into the polymer structure, offering a dual benefit of decarbonization and value creation. Recycling of polyurethanes enhances circularity but requires energy-efficient processes to compete with virgin production.

Evaluating these routes alongside electrochemical technologies is crucial to identifying scalable solutions for the polyol market. A relevant example is the "CO$_2$Exide" project (2018–2021),[86] which demonstrated co-valorization by producing ethylene and hydrogen peroxide simultaneously, enabling ethylene oxide synthesis (**Fig. 4c**).

**Industrial viewpoints: barriers for e-chemicals**

Despite great advancements, replacing conventional processes with e-chemicals faces critical barriers that hinder large-scale adoption (**Fig. 4d**). *i)* Low TRLs:[87] $CO_2$-based electrochemical methods remain at low TRLs, requiring significant advancements in catalyst stability, selectivity, current density, and energy consumption. *ii)* Market and cost barriers: The chemical industry operates with large market volumes and capital-intensive infrastructure, making the integration of new technologies challenging.[88] *iii)* Feedstock and electricity prices: The affordability of fossil-based feedstocks and the need for low electricity prices and high energy efficiency pose challenges to the cost-competitiveness of e-chemicals.[89,90] *iv)* Compatibility with renewable electricity: Continuous operation depends on reliable and affordable renewable electricity.[91,92] *v)* Policy tools: Carbon taxes and renewable origin certification can incentivize the adoption by reducing the cost disparity between e-chemicals and fossil-based processes, but their impact depends on effective implementation and complementary market conditions.[93,94]

**Emerging vs incumbent synthesis: the case study of propylene oxide**

Propylene oxide, a $200 billion market cornerstone, is vital for polyurethane and other materials, making it representative for assessing sustainable electrochemical production methods.

There are five commercially available production processes for the production of propylene oxide (**Fig. 4e**):[95,96] *i)* The chlorohydrin process (CH-PO); *ii)* The styrene monomer process (SM-PO); *iii)* The tertiary butyl alcohol/methyl tertiary butyl ether process (TBA/MTBE-PO); *iv)* The cumene process (CU-PO); *v)* The hydrogen peroxide process (HP-PO).

While traditional methods like CH-PO and SM-PO dominate the market, newer processes like HP-PO offer environmental benefits by reducing waste and improving energy efficiency (**Table 2**).

Electrochemical method demonstrates strong potential for sustainable propylene oxide production, particularly through redox-mediated systems. Early-stage studies using half-cell setups provide valuable insights into reaction mechanisms, while flow cells enable efficient operation at industrial-level current densities. For example, redox-mediated systems have achieved selectivity of 70% at 0.8 A/cm$^2$, further improving to 90% at 1.5 A/cm$^2$, showcasing the scalability and efficiency of electrochemical method for industrial applications.[53,82]

Advancing paired electrolysis hinges on both technical innovation—such as optimizing KPIs, reactor designs, and catalysts—and addressing market barriers that challenge industrial adoption. The propylene oxide case study exemplifies this balance, demonstrating how redox-mediated strategies achieve high selectivity and current density, while also highlighting challenges related to energy consumption, hazardous reagents, and cost.

**Table 2** | Comparison of propylene oxidation technologies: key process conditions, selectivity and byproducts.[97]

| Production method | Key process conditions | Selectivity to PO | Byproducts |
| --- | --- | --- | --- |
| CH-PO | Atmospheric pressure, 20–50°C | ~90% | 1,2-dichloropropane (100–150 kg), NaCl (2.1 tons per ton) or CaCl$_2$ (2.2 tons per ton), 40 tons wastewater |
| SM-PO | 120–150°C, 2–3 bar; 90–130°C, 15–60 bar | >90% | Styrene monomer (2–2.5 tons per ton of PO) |
| TBA-PO | 95–150°C, 25–55 bar | 90–95% | Tert-butanol (TBA) or methyl tertiary butyl ether (MTBE) |
| CU-PO | Atmospheric pressure, 80–120°C | >90% | Minimal byproducts |
| EC-PO | Atmospheric pressure, 25°C | >90% | Minimal byproducts, wastewater |

# Emerging strategies for efficient and sustainable valorization

## Compatibility and complete electron economy

Conventional paired electrolysis often requires compositional similarity between anolytes and catholytes, limiting its flexibility for diverse applications. Hydrogen-permeable palladium membranes (**Fig. 5a**) have emerged as an effective method for achieving full electron economy in co-valorization processes. By selectively allowing hydrogen atoms to pass through, these membranes enhance efficiency and enable the simultaneous production of valuable products, optimizing resource use. Particularly, the electrochemical oxidation of 4-methoxybenzyl alcohol to 4-methoxybenzaldehyde in water was paired with the hydrogenation of 1-hexyne to 1-hexene in an organic solvent.[98] This approach reduces membrane compatibility issues and prevents cross-contamination. Further research is needed to optimize cathode and anode valorization and validate this method in high-current flow electrolyzers.

**Redirecting products to "A(anode)+A(cathode)"**

Conventional electrolysis yields disparate products "A (anode)+B (cathode)", warranting the deployment of high-permselectivity membranes to preclude cross-contamination.[99,100]

Redirecting electrolysis to produce identical products at both electrodes ("A+A" or "B+B") simplifies downstream separation and avoids cross-contamination. For example, the oxidation of C-H bond-containing aldehydes can generate hydrogen gas at both electrodes, streamlining product recovery (**Fig. 5b, 5c**).[26]

Techniques like electro-Fenton can be adapted to generate identical high-value products at both electrodes, simplifying separation and refining steps. This strategy also reduces the risk of overoxidation by operating at lower voltages (**Fig. 5d**).[50]

**Exploiting cleaner redox-mediated oxidation**

Direct oxidation is often hindered by high energy demands, side reactions, and poor selectivity. Redox-mediated strategies, such as those used in propylene oxide production, have improved selectivity at industrial current densities up to 1 $A/cm^2$, demonstrating the potential of mediated systems for complex oxidation reactions.

However, redox-mediated catalysis faces challenges, including reliance on external agents, increased costs, and mediator stability concerns. To address these limitations, in-situ $H_2O_2$ generation presents a cleaner alternative, avoiding the corrosiveness and toxicity of halogen-based mediators while maintaining high selectivity and efficiency (**Fig. 5e**).[101]

# Conclusion

Paired electrolysis presents a promising pathway for sustainable chemical manufacturing by integrating anodic and cathodic valorization, maximizing electron economy, and enhancing energy efficiency. However, its industrial adoption faces challenges related to reactor assembly, electron transfer asymmetry, KPIs trade-offs, and process integration. Addressing these barriers requires a structured TRL-specific approach, where reactor assemblies transition from fundamental studies to industrial implementation. By strategically pairing anodic and cathodic reactions, optimizing electrocatalyst design, and developing application-specific KPIs, paired electrolysis can overcome kinetic mismatches and enhance process efficiency. The benchmark case study on propylene oxidation paired with HER or ORR demonstrates its potential against high-TRL thermocatalytic processes, providing insights into feasibility, environmental benefits, and economic trade-offs. Furthermore, hybrid integration models, along with membrane innovations, redox-mediated oxidation, and process co-valorization strategies, enable accelerated incorporation into existing industrial workflows. Moving forward, achieving industrial viability will require advancements in reactor engineering, catalyst design, downstream separation, and economic feasibility analysis. Policy support, market incentives, and cross-sector collaborations will further accelerate commercialization. By aligning electrochemical engineering principles with process scalability and sustainability goals, paired electrolysis offers a transformative approach for next-generation e-chemical manufacturing, bridging the gap between laboratory innovation and industrial reality.

# Acknowledgements

This work was partially funded by CEX2019-000910-S [MCIN/AEI/10.13039/501100011033], Fundació Cellex, Fundació Mir-Puig, and Generalitat de Catalunya through CERCA, the La Caixa Foundation [100010434, E.U. Horizon 2020 Marie Skłodowska-Curie grant agreement 847648], the European Union's Horizon 2023 research and innovation program under the Marie Sklodowska-Curie grant agreement 101150688, and REPSOL S.A.

# References

1. Hulme, M. 1.5°C and climate research after the Paris Agreement. *Nat. Clim. Change* **6**, 222–224 (2016).
2. Welsby, D., Price, J., Pye, S. & Ekins, P. Unextractable fossil fuels in a 1.5°C world. *Nature* **597**, 230–234 (2021).
3. Cushing, L. J., Li, S., Steiger, B. B. & Casey, J. A. Historical red-lining is associated with fossil fuel power plant siting and present-day inequalities in air pollutant emissions. *Nat. Energy* **8**, 52–61 (2022).
4. Renewable capacity statistics 2023. https://www.irena.org/Publications/2023/Mar/Renewable-capacity-statistics-2023, access date: 2024.11.19.
5. Kong, X. *et al.* Synthesis of hydroxylamine from air and water via a plasma-electrochemical cascade pathway. *Nat. Sustain.* **7**, 652–660 (2024).
6. Ram, R. *et al.* Water-hydroxide trapping in cobalt tungstate for proton exchange membrane water electrolysis. *Science* **384**, 1373–1380 (2024).
7. Lv, C. *et al.* Selective electrocatalytic synthesis of urea with nitrate and carbon dioxide. *Nat. Sustain.* **4**, 868–876 (2021).
8. Zhang, B. *et al.* High-valence metals improve oxygen evolution reaction performance by modulating 3d metal oxidation cycle energetics. *Nat. Catal.* **3**, 985–992 (2020).
9. Zhang, B. *et al.* Homogeneously dispersed multimetal oxygen-evolving catalysts. *Science* **352**, 333–337 (2016).
10. Vass, Á., Endrődi, B. & Janáky, C. Coupling electrochemical carbon dioxide conversion with value-added anode processes: An emerging paradigm. *Curr. Opin. Electrochem.* **25**, 100621 (2021).
11. Holst, D. E., Wang, D. J., Kim, M. J., Guzei, I. A. & Wickens, Z. K. Aziridine synthesis by coupling amines and alkenes via an electrogenerated dication. *Nature* **596**, 74–79 (2021).
12. Liu, C., Chen, F., Zhao, B.-H., Wu, Y. & Zhang, B. Electrochemical hydrogenation and oxidation of organic species involving water. *Nat. Rev. Chem.* **8**, 277–293 (2024).
13. Qiu, B. *et al.* Waste plastics upcycled for high-efficiency $H_2O_2$ production and lithium recovery via Ni-Co/carbon nanotubes composites. *Nat. Commun.* **15**, 6473 (2024).
14. Tian, C. *et al.* Progress and roadmap for electro-privileged transformations of bio-derived molecules. *Nat. Catal.* **7**, 350–360 (2024).
15. Birmingham, W. R. *et al.* Toward scalable biocatalytic conversion of 5-hydroxymethylfurfural by galactose oxidase using coordinated reaction and enzyme engineering. *Nat. Commun.* **12**, 4946 (2021).
16. Leow, W. R. *et al.* Chloride-mediated selective electrosynthesis of ethylene and propylene oxides at high current density. *Science* **368**, 1228–1233 (2020).
17. Jin, M. *et al.* Electrosynthesis of N,N-dimethylformamide from market-surplus trimethylamine coupled with hydrogen production. *Green Chem.* **25**, 5936–5944 (2023).
18. Wang, W. *et al.* Vacancy-rich $Ni(OH)_2$ drives the electrooxidation of amino C−N bonds to nitrile C≡N bonds. *Angew. Chem. Int. Edit.* **59**, 16974–16981 (2020).
19. Li, J. *et al.* Green electrosynthesis of 5,5′-azotetrazolate energetic materials plus energy-efficient hydrogen production using ruthenium single-atom catalysts. *Adv. Mater.* **34**, 2203900 (2022).
20. Sun, G.-Q. *et al.* Electrochemical reactor dictates site selectivity in N-heteroarene carboxylations. *Nature* **615**, 67–72 (2023).

21. Li, S. *et al.* Long-term continuous ammonia electrosynthesis. *Nature* **629**, 92–97 (2024).
22. Savage, T. *et al.* Machine learning-assisted discovery of flow reactor designs. *Nat. Chem. Eng.* **1**, 522–531 (2024).
23. Chen, T. *et al.* Accelerating ethanol complete electrooxidation via introducing ethylene as the precursor for the C−C bond splitting. *Angew. Chem. Int. Edit.* **62**, e202308057 (2023).
24. Leech, M. C. & Lam, K. A practical guide to electrosynthesis. *Nat. Rev. Chem.* **6**, 275–286 (2022).
25. Shen, K. *et al.* Electrochemical oxidation of methane to methanol on electrodeposited transition metal oxides. *J. Am. Chem. Soc.* **145**, 6927–6943 (2023).
26. Wang, T. *et al.* Combined anodic and cathodic hydrogen production from aldehyde oxidation and hydrogen evolution reaction. *Nat. Catal.* **5**, 66–73 (2021).
27. High-performance alkaline water electrolyzers based on Ru-perturbed Cu nanoplatelets cathode. *Nat. Commun.* **14**, 4680 (2023).
28. Wu, Z.-Y. *et al.* Non-iridium-based electrocatalyst for durable acidic oxygen evolution reaction in proton exchange membrane water electrolysis. *Nat. Mater.* **22**, 100–108 (2023).
29. Li, D. *et al.* Highly quaternized polystyrene ionomers for high performance anion exchange membrane water electrolysers. *Nat. Energy* **5**, 378–385 (2020).
30. Oener, S. Z., Foster, M. J. & Boettcher, S. W. Accelerating water dissociation in bipolar membranes and for electrocatalysis. *Science* **369**, 1099–1103 (2020).
31. Kraglund, M. R. et al. Ion-solvating membranes as a new approach towards high rate alkaline electrolyzers. *Energy Environ. Sci.* **12**, 3313–3318 (2019).
32. Ge, R. *et al.* Selective electrooxidation of biomass-derived alcohols to aldehydes in a neutral medium: Promoted water dissociation over a nickel-oxide-supported ruthenium single-atom catalyst. *Angew. Chem. Int. Edit.* **61**, e202200211 (2022).
33. Xie, K. *et al.* Eliminating the need for anodic gas separation in $CO_2$ electroreduction systems via liquid-to-liquid anodic upgrading. *Nat. Commun.* **13**, 3070 (2022).
34. Tian, C. *et al.* Paired electrosynthesis of $H_2$ and acetic acid at $A/cm^2$ current densities. *ACS Energy Lett.* **8**, 4096–4103 (2023).
35. Yan, Y. *et al.* Electrocatalytic upcycling of biomass and plastic wastes to biodegradable polymer monomers and hydrogen fuel at high current densities. *J. Am. Chem. Soc.* **145**, 6144–6155 (2023).
36. Lou, X. *et al.* Highly efficient and low cost $SPEEK/TiO_2$ nanocomposite membrane for vanadium redox flow battery. *J. Nanosci. Nanotechnol.* **19**, 2247–2252 (2019).
37. Ye, J. *et al.* Hybrid membranes dispersed with superhydrophilic $TiO_2$ nanotubes toward ultra-stable and high-performance vanadium redox flow batteries. *Adv. Energy Mater.* **10**, 1904041 (2020).
38. Staszak-Jirkovský, J. *et al.* Design of active and stable Co–Mo–$S_x$ chalcogels as pH-universal catalysts for the hydrogen evolution reaction. *Nat. Mater.* **15**, 197–203 (2016).
39. Mahmood, J. *et al.* An efficient and pH-universal ruthenium-based catalyst for the hydrogen evolution reaction. *Nat. Nanotech.* **12**, 441–446 (2017).
40. Kosmala, T. *et al.* Operando visualization of the hydrogen evolution reaction with atomic-scale precision at different metal–graphene interfaces. *Nat. Catal.* **4**, 850–859 (2021).
41. Lu, Z. *et al.* High-efficiency oxygen reduction to hydrogen peroxide catalysed by oxidized carbon materials. *Nat. Catal.* **1**, 156–162 (2018).
42. Santoro, C., Bollella, P., Erable, B., Atanassov, P. & Pant, D. Oxygen reduction reaction

electrocatalysis in neutral media for bioelectrochemical systems. *Nat. Catal.* **5**, 473–484 (2022).

43. Wang, X. *et al.* Efficient electrically powered $CO_2$-to-ethanol via suppression of deoxygenation. *Nat. Energy* **5**, 478–486 (2020).

44. Ozden, A. *et al.* Carbon-efficient carbon dioxide electrolysers. *Nat. Sustain.* **5**, 563–573 (2022).

45. Fan, R. *et al.* Ultrastable electrocatalytic seawater splitting at ampere-level current density. *Nat. Sustain.* **7**, 158–167 (2024).

46. Zhao, S. *et al.* Selective electrosynthesis of chlorine disinfectants from seawater. *Nat. Sustain.* **7**, 148–157 (2024).

47. Zhou, H. *et al.* Scalable electrosynthesis of commodity chemicals from biomass by suppressing non-Faradaic transformations. *Nat. Commun.* **14**, 5621 (2023).

48. Verma, S., Lu, S. & Kenis, P. J. A. Co-electrolysis of $CO_2$ and glycerol as a pathway to carbon chemicals with improved technoeconomics due to low electricity consumption. *Nat. Energy* **4**, 466–474 (2019).

49. Patel, P. *et al.* Technoeconomic and life-cycle assessment for electrocatalytic production of furandicarboxylic acid. *ACS Sustain. Chem. Eng.* **10**, 4206–4217 (2022).

50. Sheng, H. *et al.* Linear paired electrochemical valorization of glycerol enabled by the electro-Fenton process using a stable $NiSe_2$ cathode. *Nat. Catal.* **5**, 716–725 (2022).

51. Zuo, K. Electrified water treatment: Fundamentals and roles of electrode materials. *Nat. Rev. Mater.*

52. Na, J. *et al.* General technoeconomic analysis for electrochemical coproduction coupling carbon dioxide reduction with organic oxidation. *Nat. Commun.* **10**, 5193 (2019).

53. Leow, W. R. *et al.* Chloride-mediated selective electrosynthesis of ethylene and propylene oxides at high current density. *Science* **368**, 1228–1233 (2020).

54. Li, Z. *et al.* Iridium single-atom catalyst on nitrogen-doped carbon for formic acid oxidation synthesized using a general host–guest strategy. *Nat. Chem.* **12**, 764–772 (2020).

55. Jeyaseelan, A., Naushad, M., Ahamad, T. & Viswanathan, N. Fabrication of amino functionalized benzene-1,4-dicarboxylic acid facilitated cerium based metal organic frameworks for efficient removal of fluoride from water environment. *Environ. Sci.: Water Res. Technol.* **7**, 384–395 (2021).

56. Hintz, H. A. & Sevov, C. S. Catalyst-controlled functionalization of carboxylic acids by electrooxidation of self-assembled carboxyl monolayers. *Nat. Commun.* **13**, 1319 (2022).

57. Proost, J. State-of-the art CAPEX data for water electrolysers, and their impact on renewable hydrogen price settings. *Int. J. Hydrogen Energ.* **44**, 4406–4413 (2019).

58. Asian ethylene, propylene prices under heavy strain from oversupply | ChemOrbis. https://www.chemorbis.com/en/plastics-news/Asian-ethylene-propylene-prices-under-heavy-strain-from-oversupply/2021/12/15/832094&isflashhaber=true#reportH, access date: 2024.11.19.

59. Propylene oxide: market demand, growth, share & outlook 2023-2035. https://www.researchnester.com/reports/propylene-oxide-market/2804, access data: 2024.11.19.

60. Global ethylene market analysis report 2023: Plant capacity, production, operating efficiency, demand/supply, end-users, sales channel, regional demand, foreign trade, company shares, 2015-2035. *Yahoo Finance* https://finance.yahoo.com/news/global-ethylene-market-analysis-report-123300287.html, access date: 2024.11.19.

61. Kim, H., Baek, S. & Won, W. Integrative technical, economic, and environmental sustainability analysis for the development process of biomass-derived 2,5-furandicarboxylic acid. *Renew. Sust.*

*Energ. Rev.* **157**, 112059 (2022).

62. Techno-economic analysis and life cycle assessment of the production of biodegradable polyaliphatic–polyaromatic polyesters. *ACS Sustain. Chem. Eng.* **12**, 9156–9167 (2024).

63. China propylene oxide spot price, China propylene oxide daily prices provided by SunSirs, China Commodity Data Group. https://www.sunsirs.com/uk/prodetail-438.html, access date: 2024.11.19.

64. Ethylene oxide prices | historical & forecast | intratec.us. https://www.intratec.us/chemical-markets/ethylene-oxide-price, access date: 2024.11.19.

65. Garedew, M. *et al.* Greener routes to biomass waste valorization: lignin transformation through electrocatalysis for renewable chemicals and fuels production. *ChemSusChem* **13**, 4214–4237 (2020).

66. Ke, J. *et al.* Dynamically reversible interconversion of molecular catalysts for efficient electrooxidation of propylene into propylene glycol. *J. Am. Chem. Soc.* **145**, 9104–9111 (2023).

67. Ke, J. *et al.* Facet-dependent electrooxidation of propylene into propylene oxide over $Ag_3PO_4$ crystals. *Nat. Commun.* **13**, 932 (2022).

68. Chung, M. *et al.* Direct propylene epoxidation via water activation over Pd-Pt electrocatalysts. *Science* **383**, 49–55 (2024).

69. von Münchow, T., Dana, S., Xu, Y., Yuan, B. & Ackermann, L. Enantioselective electrochemical cobalt-catalyzed aryl C–H activation reactions. *Science* **379**, 1036–1042 (2023).

70. Mao, Q. *et al.* Atomically dispersed Cu coordinated Rh metallene arrays for simultaneously electrochemical aniline synthesis and biomass upgrading. *Nat. Commun.* **14**, 5679 (2023).

71. Li, T. *et al.* Cobalt-catalyzed atroposelective C−H activation/annulation to access N−N axially chiral frameworks. *Nat. Commun.* **14**, 5271 (2023).

72. Liu, F., Gao, X., Shi, R., Tse, E. C. M. & Chen, Y. A general electrochemical strategy for upcycling polyester plastics into added-value chemicals by a $CuCo_2O_4$ catalyst. *Green Chem.* **24**, 6571–6577 (2022).

73. Liu, K. *et al.* Selective electrocatalytic reforming of PET-derived ethylene glycol to formate with a Faraday efficiency of 93.2% at industrial-level current densities. *Chem. Eng. J.* **473**, 145292 (2023).

74. Wang, X.-H. *et al.* Ultra-thin $CoNi_{0.2}P$ nanosheets for plastics and biomass participated hybrid water electrolysis. *Chem. Eng. J.* **465**, 142938 (2023).

75. Tan, R. *et al.* Hydrophilic microporous membranes for selective ion separation and flow-battery energy storage. *Nat. Mater.* **19**, 195–202 (2020).

76. Zhou, S. *et al.* Asymmetric pore windows in MOF membranes for natural gas valorization. *Nature* **606**, 706–712 (2022).

77. Wi, T.-U. *et al.* Upgrading carbon monoxide to bioplastics via integrated electrochemical reduction and biosynthesis. *Nat. Synth.* **3**, 1392–1403 (2024).

78. Zheng, T. *et al.* Upcycling $CO_2$ into energy-rich long-chain compounds via electrochemical and metabolic engineering. *Nat. Catal.* **5**, 388–396 (2022).

79. Zheng, T. *et al.* Copper-catalysed exclusive $CO_2$ to pure formic acid conversion via single-atom alloying. *Nat. Nanotechnol.* **16**, 1386–1393 (2021).

80. Xia, C., Xia, Y., Zhu, P., Fan, L. & Wang, H. Direct electrosynthesis of pure aqueous $H_2O_2$ solutions up to 20% by weight using a solid electrolyte. *Science* **366**, 226–231 (2019).

81. Hao, Y. *et al.* Methanol upgrading coupled with hydrogen product at large current density

promoted by strong interfacial interactions. *Energy Environ. Sci.* **16**, 1100–1110 (2023).

82. Li, Y. *et al.* Redox-mediated electrosynthesis of ethylene oxide from $CO_2$ and water. *Nat. Catal.* **5**, 185–192 (2022).

83. Shi, R. *et al.* Electrocatalytic reforming of waste plastics into high value-added chemicals and hydrogen fuel. *Chem. Comm.* **57**, 12595–12598 (2021).

84. Liu, F. *et al.* Concerted and selective electrooxidation of polyethylene-terephthalate-derived alcohol to glycolic acid at an industry-level current density over a Pd−Ni(OH)$_2$ catalyst. *Angew. Chem. Int. Edit.* **62**, e202300094 (2023).

85. Chen, Z.-J. *et al.* Acidic enol electrooxidation-coupled hydrogen production with ampere-level current density. *Nat. Commun.* **14**, 4210 (2023).

86. Polymers with intrinsic microporosity (PIMs) for targeted $CO_2$ reduction to ethylene | H2020. *CORDIS | European Commission* https://cordis.europa.eu/project/id/768789/reporting, access date: 2024.11.19.

87. Repsol will be a net zero emissions company by 2050, https://www.repsol.com/en/press-room/press-releases/2019/repsol-will-be-a-net-zero-emissions-company-by-2050/index.cshtml, access date: 2024.11.19.

88. De Luna, P. *et al.* What would it take for renewably powered electrosynthesis to displace petrochemical processes? *Science* **364**, eaav3506 (2019).

89. Lux research predicts which chemical manufacturing platforms will be electrified by 2050. https://coatingsworld.com/contents/view_experts-opinion/2020-04-07/lux-research-predicts-which-chemical-manufacturing-platforms-will-be-electrified-by-2050/, access date: 2024.11.19.

90. Glenk, G. & Reichelstein, S. Economics of converting renewable power to hydrogen. *Nat. Energy* **4**, 216–222 (2019).

91. Headley, A. J. & Copp, D. A. Energy storage sizing for grid compatibility of intermittent renewable resources: A California case study. *Energy* **198**, 117310 (2020).

92. Diesendorf, M. & Elliston, B. The feasibility of 100% renewable electricity systems: A response to critics. *Renew. Sust. Energ. Rev.* **93**, 318–330 (2018).

93. Muñoz, M., Oschmann, V. & David Tàbara, J. Harmonization of renewable electricity feed-in laws in the European Union. *Energy Policy* **35**, 3104–3114 (2007).

94. Elkins, P. & Baker, T. Carbon taxes and carbon emissions trading. *Journal of Economic Surveys* **15**, 325–376 (2001).

95. Soh, J. Is the global propylene market on the rise? *Chemical Market Analytics by OPIS,* https://chemicalmarketanalytics.com/blog/european_polyurethane_feedstocks_market/, access date: 2024.11.19.

96. Propylene oxide (2017 Program). *NexantECA* https://www.nexanteca.com/reports/propylene-oxide-2017-program, access date: 2024.11.19.

97. Using online chemical analysis to optimize propylene oxide production. https://www.metrohm.com/es_es/applications/whitepaper/wp-048.html, access date: 2024.11.19

98. Sherbo, R. S., Delima, R. S., Chiykowski, V. A., MacLeod, B. P. & Berlinguette, C. P. Complete electron economy by pairing electrolysis with hydrogenation. *Nat. Catal.* **1**, 501–507 (2018).

99. Chandra, A., Das, S., Nir, O. & Gilron, J. A facile tuning of polyelectrolyte cross-linking chemistry for improved permselectivity and stability of a monovalent cation exchange membrane. *Desalination* **586**, 117874 (2024).

100. Wang, X. *et al.* Solvent-vapor-triggered crystallization of a ZIF-90 membrane with versatile

separation properties towards light hydrocarbons. *Chem. Eng. J.* **496**, 153737 (2024).

101. Kim, J. *et al.* Electro-assisted methane oxidation to formic acid via in-situ cathodically generated $H_2O_2$ under ambient conditions. *Nat. Commun.* **14**, 4704 (2023).

102. Wang, C. *et al.* A novel electrode for value-generating anode reactions in water electrolyzers at Industrial Current Densities. *Angew. Chem. Int. Ed.* **135**, (2023).

103. Zhu, Y. *et al.* Constructing Ru-O-TM bridge in NiFe-LDH enables high current hydrazine-assisted $H_2$ Production. *Adv. Mater.* **36**, 2401694 (2024).

104. Li, J. *et al.* Green electrosynthesis of 3,3'-diamino-4,4'-azofurazan energetic materials coupled with energy-efficient hydrogen production over Pt-based catalysts. *Nat. Commun.* **14**, 8146 (2023).

105. Armstrong, D. A. *et al.* Standard electrode potentials involving radicals in aqueous solution: inorganic radicals (IUPAC Technical Report). *Pure and Applied Chemistry* **87**, 1139–1150 (2015).

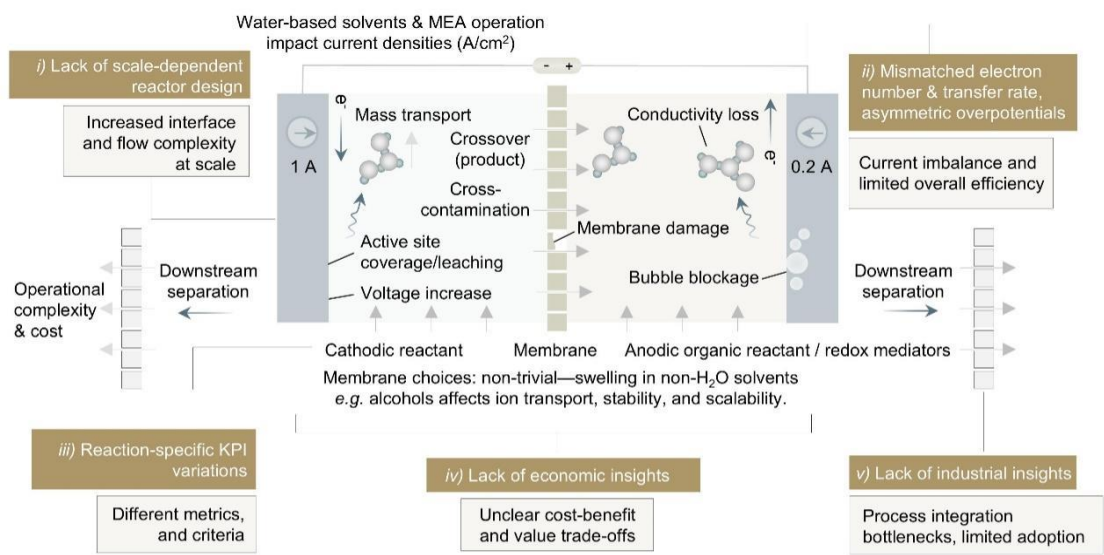

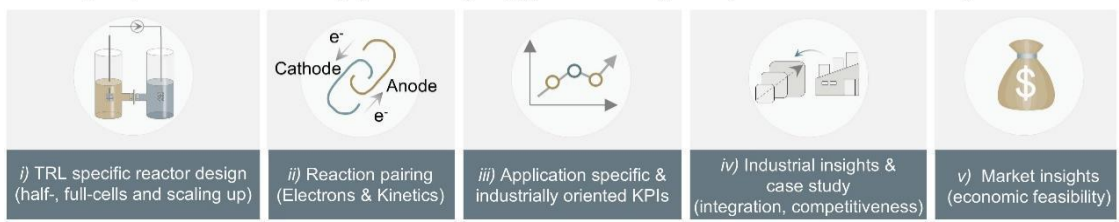

**Fig. 1 | Key challenges in industrial paired electrolysis and a holistic strategy to address them. a.** Industrial paired electrolysis faces challenges such as mismatched electron kinetics, mass transport limitations, scale-dependent reactor complexity, and unclear economic viability. **b.** This work offers a vision of these challenges and opportunities over a system-level framework to optimize reaction pairing, reactor design, and techno-economic integration, enabling practical large-scale implementation.

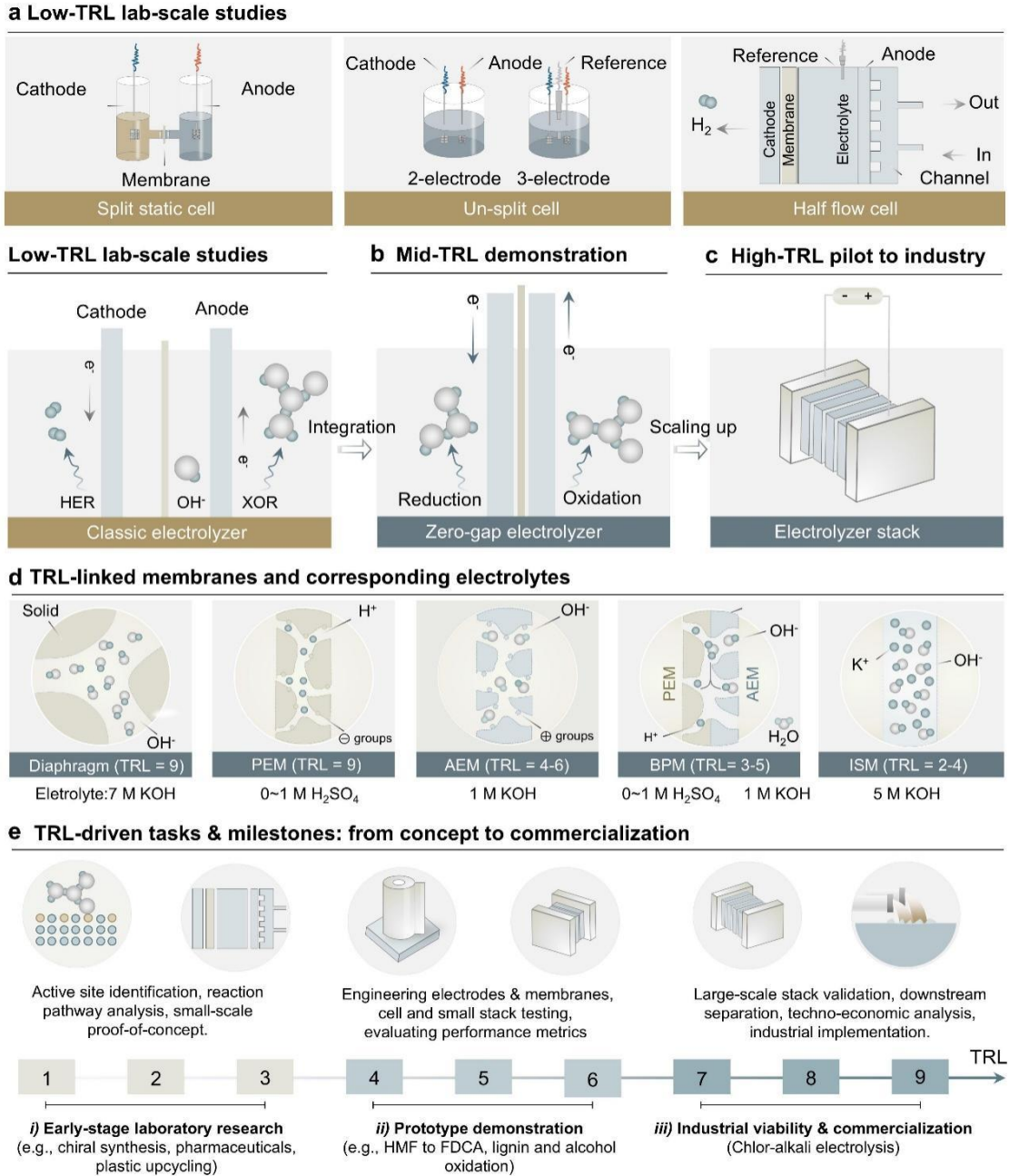

**Fig. 2 | TRL-specific development pathway for paired electrolysis, from laboratory studies to industrial deployment. a,** Low-TRL lab-scale studies involve split static cells, un-split two- and three-electrode configurations, and half-flow cells. **b,** Mid-TRL demonstration transitions to zero-gap electrolyzers. **c,** High-TRL pilot to industry involves scaling up to industrial electrolyzer stacks. **d,** TRL-linked membranes and corresponding electrolytes illustrate how membrane selection evolves with TRL, from diaphragm and PEM systems to AEM, BPM, and ISM, each requiring tailored electrolyte environments. **e,** TRL-driven tasks & milestones outline the pathway from early-stage laboratory research, through prototype demonstration, to industrial validation and commercialization.

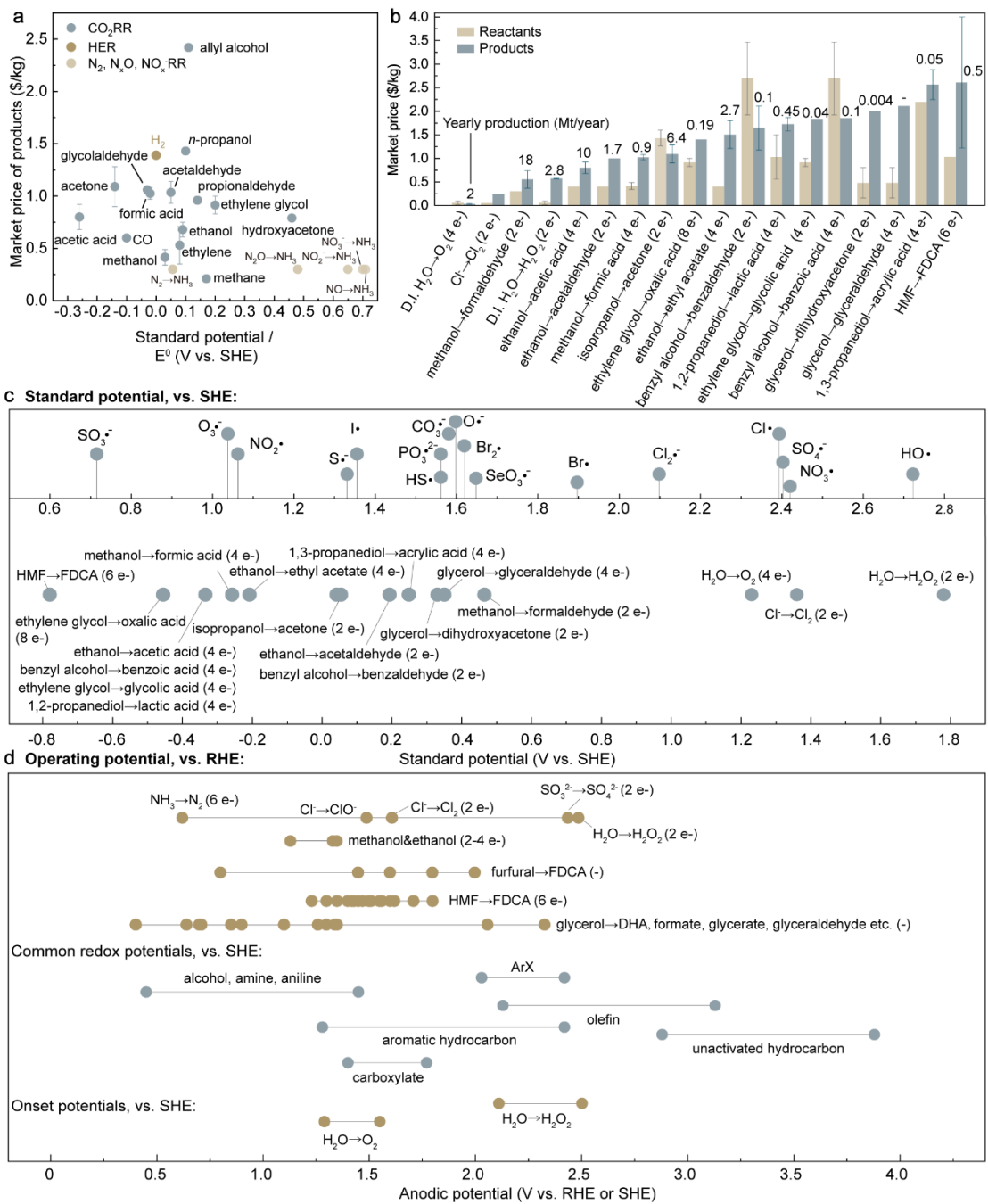

**Fig. 3** | Market and thermodynamic analysis. **a,** Market price corresponding to standard potential of electrochemical reduction reactions.[10, 52] **b,** Market price of reactants and products produced by electrochemical oxidation reactions.[10, 52] **c,** Standard potential of representative oxidation reactions for the formation of intermediates,[51, 105] and electrochemical valorization.[52] **d,** Anodic operating potential of practical oxidation reactions in half cells (See **Table 1** in Supplementary Information for details).

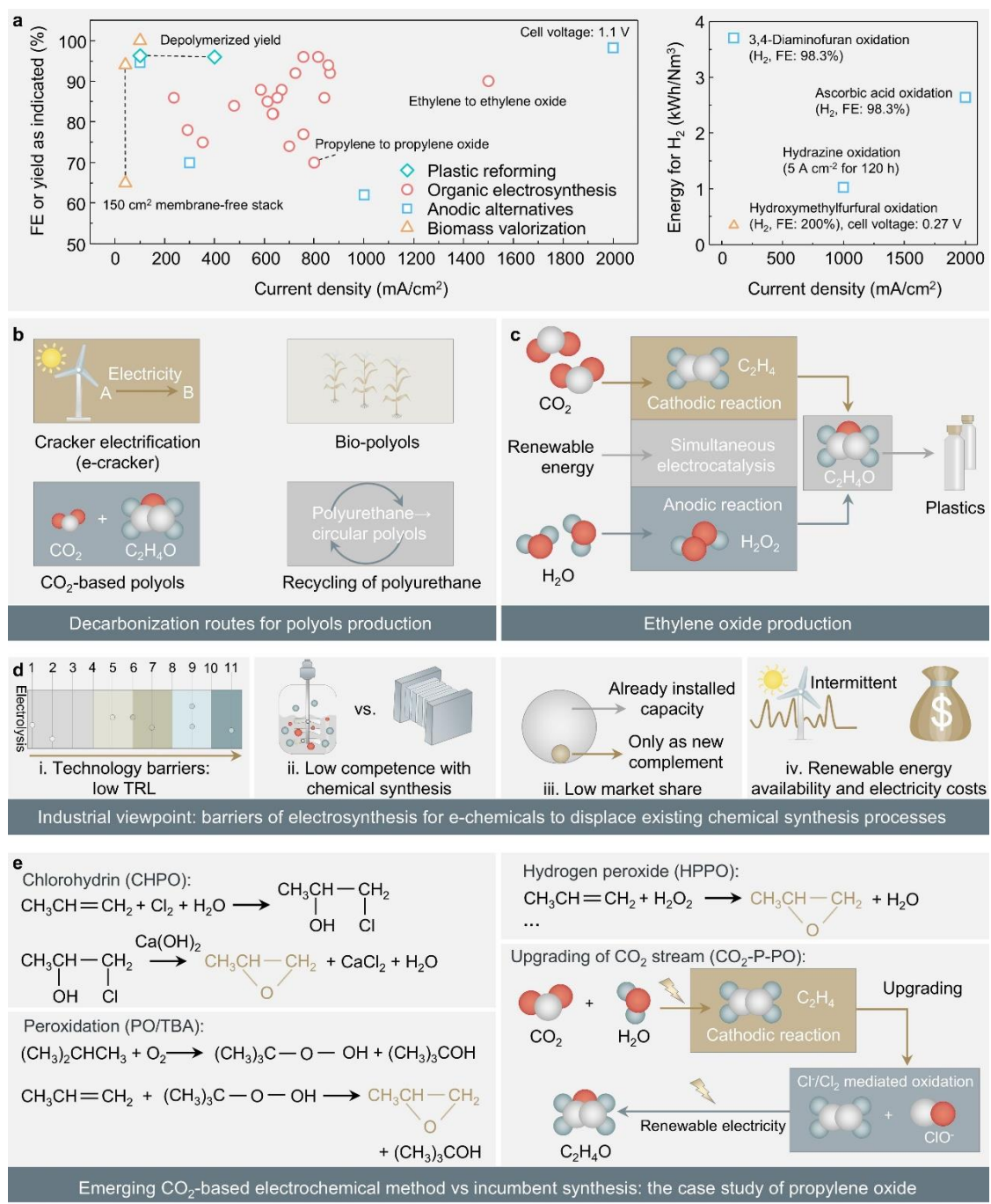

**Fig. 4 | Performance metrics and the case study: KPIs, industrial examples and viewpoints**. **a,** KPIs, *e.g.,* current density, Faraday efficiency (FE) and stability for high-current plastic reforming,[83,84] biomass valorization,[26,35] organic electrosynthesis,[53,82,102] and other anodic alternatives.[85,103,104] **b, c,** Decarbonization routes for polyols production and co-electrolysis for ethylene oxide production. **d,** Barriers of electrosynthesis for e-chemicals to displace existing chemical synthesis processes. **e.** Comparison among traditional and $CO_2$-based electrochemical methods to produce propylene oxide.

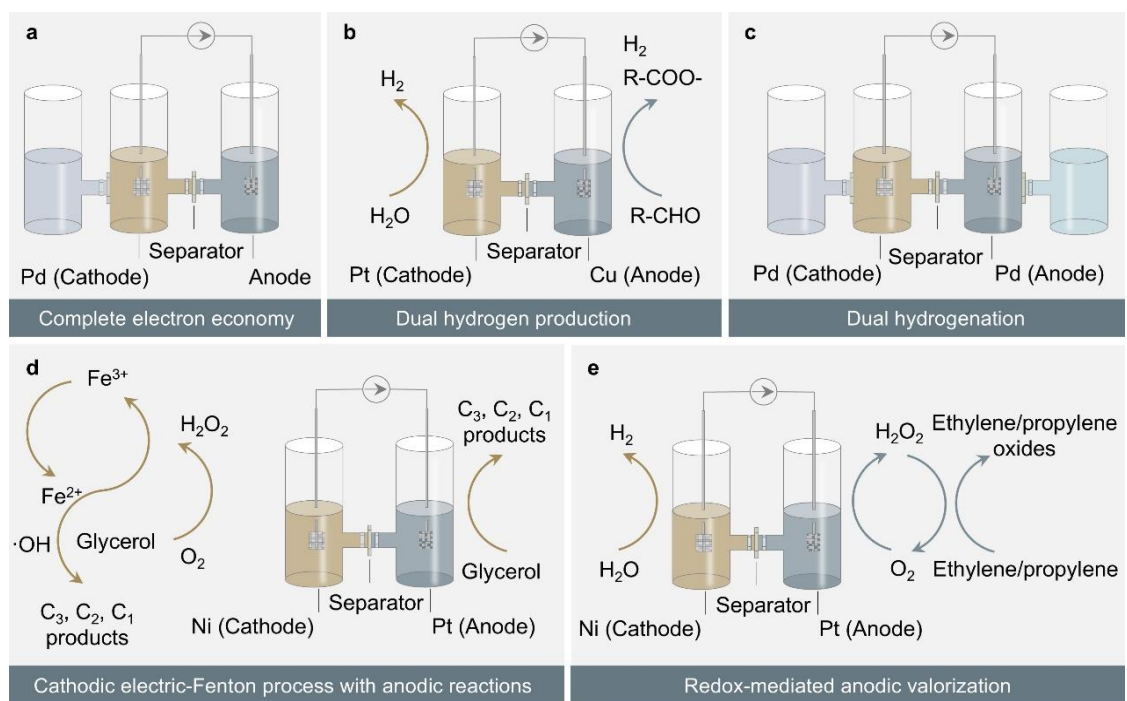

**Fig. 5 | Emerging concepts encouraging to paired electrolysis. a,** Achieving complete electron economy through innovative reactor assemblies. **b,** Dual hydrogen production via controlled oxidation. **c,** Dual hydrogenation via controlled oxidation and hydrogen conductive palladium electrodes. Combining anodic decontamination techniques with valorization strategies on both electrodes: **d,** Electric-Fentone process, **e.** redox-mediated catalysis.